\documentclass[apjl]{emulateapj}

\usepackage{txfonts}

  \SetSymbolFont{symbols}{bold}{OMS}{cmsy}{b}{n}
  \DeclareSymbolFont{bmisymbols}{OML}{cmm}{b}{it}
  \DeclareMathSymbol{\balpha}{0}{bmisymbols}{"0B}
  \DeclareMathSymbol{\bbeta}{0}{bmisymbols}{"0C}
  \DeclareMathSymbol{\bgamma}{0}{bmisymbols}{"0D}
  \DeclareMathSymbol{\bdelta}{0}{bmisymbols}{"0E}
  \DeclareMathSymbol{\bepsilon}{0}{bmisymbols}{"0F}
  \DeclareMathSymbol{\bzeta}{0}{bmisymbols}{"10}
  \DeclareMathSymbol{\boldeta}{0}{bmisymbols}{"11}
  \DeclareMathSymbol{\btheta}{0}{bmisymbols}{"12}
  \DeclareMathSymbol{\biota}{0}{bmisymbols}{"13}
  \DeclareMathSymbol{\bkappa}{0}{bmisymbols}{"14}
  \DeclareMathSymbol{\blambda}{0}{bmisymbols}{"15}
  \DeclareMathSymbol{\bmu}{0}{bmisymbols}{"16}
  \DeclareMathSymbol{\bnu}{0}{bmisymbols}{"17}
  \DeclareMathSymbol{\bxi}{0}{bmisymbols}{"18}
  \DeclareMathSymbol{\bpi}{0}{bmisymbols}{"19}
  \DeclareMathSymbol{\brho}{0}{bmisymbols}{"1A}
  \DeclareMathSymbol{\bsigma}{0}{bmisymbols}{"1B}
  \DeclareMathSymbol{\btau}{0}{bmisymbols}{"1C}
  \DeclareMathSymbol{\bupsilon}{0}{bmisymbols}{"1D}
  \DeclareMathSymbol{\bphi}{0}{bmisymbols}{"1E}
  \DeclareMathSymbol{\bchi}{0}{bmisymbols}{"1F}
  \DeclareMathSymbol{\bpsi}{0}{bmisymbols}{"20}
  \DeclareMathSymbol{\bomega}{0}{bmisymbols}{"21}
  \DeclareMathSymbol{\bvarepsilon}{0}{bmisymbols}{"22}
  \DeclareMathSymbol{\bvartheta}{0}{bmisymbols}{"23}
  \DeclareMathSymbol{\bvarpi}{0}{bmisymbols}{"24}
  \DeclareMathSymbol{\bvarrho}{0}{bmisymbols}{"25}
  \DeclareMathSymbol{\bvarsigma}{0}{bmisymbols}{"26}
  \DeclareMathSymbol{\bvarphi}{0}{bmisymbols}{"27}

\newcommand{\mathbfit}[1]{\textbf{\textit{#1}}}

\newcommand{\Mdot}{\dot{M}}

\newcommand{\elct}{\mathrm{e}}
\newcommand{\rmn}{\mathrm}
\newcommand{\dd}{\mathrm{d}}
\newcommand{\vecbf}{\mathbfit}
\newcommand{\vel}{\upsilon}
\newcommand{\bvel}{\bupsilon}
\newcommand{\hsev}{h_{70}}
\newcommand{\asz}{A_{\mathrm{SZ}}}
\newcommand{\atsz}{A_{\mathrm{tSZ}}}
\newcommand{\aksz}{A_{\mathrm{kSZ}}}

\slugcomment{Submitted to ApJ}

\shorttitle{Simulations of SZ with AGN feedback}
\shortauthors{Battaglia et al.}

\begin{document}

\title{Simulations of the Sunyaev-Zel'dovich Power Spectrum with AGN Feedback}

\author{N. Battaglia\altaffilmark{1,2}, J. R. Bond\altaffilmark{2}, C. Pfrommer\altaffilmark{2},  J. L. Sievers\altaffilmark{2} and D. Sijacki\altaffilmark{3}}

\altaffiltext{1}{ Department of Astronomy and Astrophysics, University of Toronto, 50 St George , Toronto ON, M5S 3H4.}
\altaffiltext{2}{Canadian Institute for Theoretical Astrophysics, 60 St George , Toronto ON, M5S 3H8}
\altaffiltext{3}{ Kavli Institute for Cosmology, Cambridge and Institute of Astronomy, Madingley Road, Cambridge, CB3 0HA, United Kingdom}

\begin{abstract}
  We explore how radiative cooling, supernova feedback, cosmic rays
  and a new model of the energetic feedback from active galactic
  nuclei (AGN) affect the thermal and kinetic Sunyaev-Zel'dovich (SZ)
  power spectra. To do this, we use a suite of hydrodynamical
  TreePM-SPH simulations of the cosmic web in large periodic boxes and
  tailored higher resolution simulations of individual galaxy
  clusters. Our AGN feedback simulations match the recent universal
  pressure profile and cluster mass scaling relations of the REXCESS
  X-ray cluster sample better than previous analytical or numerical
  approaches. For multipoles $\ell\lesssim 2000$, our power spectra
  with and without enhanced feedback are similar, suggesting
  theoretical uncertainties over that range are relatively small,
  although current analytic and semi-analytic approaches overestimate
  this SZ power. We find the power at high $2000-10000$ multipoles
  which ACT and SPT probe is sensitive to the feedback prescription,
  hence can constrain the theory of intracluster gas, in particular
  for the highly uncertain redshifts $>0.8$. The apparent tension
  between $\sigma_8$ from primary cosmic microwave background power
  and from analytic SZ spectra inferred using ACT and SPT data is
  lessened with our AGN feedback spectra.
\end{abstract}


\section{SZ Power Templates and the Overcooling Problem}
When CMB photons are Compton-scattered by hot electrons, they gain
energy, giving a spectral decrement in thermodynamic temperature below
$\nu \approx 220$~GHz, and an excess above
\citep{1970Ap&SS...7....3S}. The high electron pressures in the
intracluster medium (ICM) result in cluster gas dominating the effect.
The integrated signal is proportional to the cluster thermal energy
and the differential signal probes the pressure profile. The SZ sky is
therefore an effective tool for constraining the internal physics of
clusters and cosmic parameters associated with the growth of
structure, in particular the {\it rms} amplitude of the (linear)
density power spectrum on cluster-mass scales $\sigma_8$
\citep[e.g.,][]{1999PhR...310...97B, 2002ARA&A..40..643C}. Identifying
clusters through blind SZ surveys and measuring the SZ power spectrum
have been long term goals in CMB research, and are reaching fruition
through the South Pole Telescope, SPT \citep{2009arXiv0912.4317L} and
Atacama Cosmology Telescope, ACT \citep{2010arXiv1001.2934T}
experiments. The ability to determine cosmological parameters from
these SZ measurements is limited by the systematic uncertainty in
theoretical modelling of the underlying cluster physics and hence of
the SZ power spectrum. The power contribution due to the kinetic SZ
(kSZ) effect that arises from ionized gas motions with respect to the
CMB rest frame adds additional uncertainty.

There are two main approaches to theoretical computations of the
thermal SZ (tSZ) power spectrum: from hydrodynamical simulations of SZ
sky maps or from semi-analytical estimates
\citep[][B0205]{2002ASPC..257...15B,2005ApJ...626...12B}. Large
cosmological simulations providing a gastrophysical solution to the
pressure distribution should include effects of non-virialized
motions, accretion shocks, and deviations from spherical
symmetry. Averaging over many realizations of synthetic SZ sky
projections yields the power spectrum and its variance \citep[e.g.,
B0205;]
[]{2000MNRAS.317...37D,2001ApJ...549..681S,2001PhRvD..63f3001S,2006MNRAS.370.1309S}. In
conjunction with primary anisotropy signals and extragalactic source
models, the SZ power spectrum has been used as a template with
variable amplitude $A_{\rmn{SZ}}$ for extracting cosmological
parameters by the Cosmic Background Imager (CBI) team
\citep[B0205;][]{2009arXiv0901.4540S} and the ACBAR team
\citep{2003ApJ...599..773G,2009ApJ...694.1200R}. $A_{\rmn{SZ}}$ was
used to estimate a $\sigma_{\mathrm{8, SZ}}\propto A_{\rmn{SZ}}^{1/7}$
as a way to encode tension between the SZ-determined value and the
(lower) $\sigma_8$ obtained from the primary anisotropy signal. The
CBI team also has included an analytic model
\citep[][KS]{2002MNRAS.336.1256K} which was also the one adopted by
the WMAP team \citep{2007ApJS..170..377S}. The KS template yielded a
lower value for $\sigma_{\mathrm{8, SZ}}$ than that obtained with the
simulation template, by $\sim 10\%$. The KS model assumes a universal
ICM pressure profile in hydrostatic equilibrium with a polytropic
(constant $\Gamma$) equation of state. The power spectrum is then
obtained using an analytic fit to `halo model' abundances. So far the
SPT and ACT have only used the KS template and a related semi-analytic
one \citep{2005ApJ...634..964O, 2009ApJ...700..989B}. This model
\citep[][S10]{2010ApJ...709..920S} allows map generation by painting
dark matter halos in N-body simulations with gas. It expands on KS by
calculating the gravitational potential from the DM particles,
includes an effective infall pressure, adds simplified models for star
formation, non-thermal pressure support and energy feedback which are
calibrated to observations. Using these templates, the SPT team
derived a $\sigma_{\mathrm{8, SZ}}$ lower than the primary anisotropy
$\sigma_8$ \citep[e.g., WMAP7,][]{2010arXiv1001.4635L}.

Current simulations with {\em only} radiative cooling and supernova
feedback excessively over-cool cluster centers
\citep[e.g.][]{2000ApJ...536..623L}, leading to too many stars in the
core, an unphysical rearrangement of the thermal and hydrodynamic
structure, and problems when compared to observations, in particular
for the entropy and pressure profiles. The average ICM pressure
profile found through X-ray observations of a sample of nearby galaxy
clusters \citep{2009arXiv0910.1234A} is inconsistent with
adaptive-mesh cluster simulations \citep{2007ApJ...668....1N}, as well
as the KS analytic model \citep{2010arXiv1001.4538K}.  Pre-heating
\citep[e.g.][]{2001ApJ...555..597B} and AGN feedback
\citep[e.g.][]{2007MNRAS.380..877S, 2008MNRAS.387.1403S,
2008ApJ...687L..53P} help solve the over-cooling problem and improve
agreement with observed cluster properties.

Previously, an analytical model by \citet{2005ApJ...634...90R} has
explored the effects of effervescent heating on the SZ power spectrum
and \citet{2007MNRAS.382.1697H} use a semi-analytical model to
calculate how an entropy floor affects the SZ power spectrum. There
have been several simulations on galaxy and group scales that have studied
how `quasar' feedback impacts the total SZ decrement
\citep{2006ApJ...653...86T, 2008ApJ...678..674S, 2008MNRAS.389...34B,
2008MNRAS.390..535C}. In this work we explore whether AGN feedback
incorporated into hydrodynamical simulations of structure formation
can suppress the over-cooling problem and resolve the current
inconsistency between theoretical predictions and observations of the
SZ power spectrum and X-ray pressure profile.

\section{Modeled physics in our simulations}

\subsection{Cosmological simulations}

We pursue two complementary approaches using smoothed particle
hydrodynamic (SPH) simulations: large-scale periodic boxes provide us
with the necessary statistics and volume to measure the SZ power
spectrum; individual cluster computations allow us to address
over-cooling at higher resolution and compare our AGN feedback
prescription with previous models. We used a modified version of the
GADGET-2 \citep{2005MNRAS.364.1105S} code.  Our sequence of periodic
boxes had sizes $100, 165, 330\,h^{-1}\,\rmn{Mpc}$. The latter two
used $N_\rmn{DM}=N_{\mathrm{gas}} = 256^3$ and $512^3$, maintaining
the same gas particle mass $m_{\mathrm{gas}} = 3.2\times 10^9\,
h^{-1}\,\mathrm{M}_{\sun}$, DM particle mass $m_{\mathrm{DM}} = 1.54\times 10^{10}\,
h^{-1}\,\mathrm{M}_{\sun}$ and a minimum gravitational smoothing
length $\varepsilon_\rmn{s}=20\, h^{-1}\,$kpc; our SPH densities were
computed with 32 neighbours. For our standard calculations, we adopt a
tilted $\Lambda$CDM cosmology, with total matter density (in units of
the critical) $\Omega_{\mathrm{m}}$= $\Omega_{\mathrm{DM}}$ +
$\Omega_{\mathrm{b}}$ = 0.25, baryon density $\Omega_{\mathrm{b}}$ =
0.043, cosmological constant $\Omega_{\Lambda}$ = 0.75, Hubble
parameter $h$ = 0.72 in units of $100 \mbox{ km s}^{-1} \mbox{
Mpc}^{-1}$, spectral index of the primordial power-spectrum $n_{\rmn{s}}$ =
0.96 and $\sigma_8$ = 0.8. For the `zoomed' cases
\citep{1993ApJ...412..455K}, we repeatedly simulated the cluster
`g676'(with the high resolution $m_{\mathrm{gas}} = 1.7\times 10^8\,
h^{-1}\,\mathrm{M}_{\sun}$, $m_{\mathrm{DM}} = 1.13\times 10^9\,
h^{-1}\,\mathrm{M}_{\sun}$ and $\varepsilon_\rmn{s} = 5\, h^{-1}$~kpc,
using 48 neighbours to compute SPH densities, as in
\citealt{2007MNRAS.378..385P}).

We show results for three variants of gas heating: (1) the classic
non-radiative `adiabatic' case with only formation {\it shock
heating}; (2) an extended {\it radiative cooling} case with star
formation, supernova (SN) feedback and cosmic rays (CRs) from
structure formation shocks; (3) {\it AGN feedback} in addition to
radiative cooling, star formation, and SN feedback. Radiative cooling
and heating were computed assuming an optically thin gas of a pure
hydrogen and helium primordial composition in a time-dependent,
spatially uniform ultraviolet background. Star formation and
supernovae feedback were modelled using the hybrid multiphase model
for the interstellar medium of \citet{2003MNRAS.339..289S}. The CR
population is modelled as a relativistic population of protons
described by an isotropic power-law distribution function in momentum
space with a spectral index of $\alpha=2.3$, following
\citet{2007A&A...473...41E}. With those parameters, the CR pressure
modifies the SZ effect at most at the percent level and causes a
reduction of the resulting integrated Compton-$y$ parameter
\citep{2007MNRAS.378..385P}.

\subsection{AGN feedback model}

Current state-of-the-art cosmological simulations are still unable to
span the large range of scales needed to resolve black hole accretion.
Hence a compromise treatment for AGN feedback is needed. For example,
\citet{2007MNRAS.380..877S} and \citet{2009MNRAS.398...53B} adopted
estimates of black hole accretion rates based on the
Bondi-Hoyle-Lyttleton formula \citep{1944MNRAS.104..273B}. Here we
introduce a sub-grid AGN feedback prescription for clusters that
allows for lower resolution still and hence can be applied to
large-scale structure simulations.  We couple the black hole accretion
rate to the global star formation rate (SFR) of the cluster, as
suggested by \citet{2005ApJ...630..167T} using the following
arguments. The typical black hole accretion rates and masses for the
inner gravitationally stable AGN disks (of size $\la 1\mathrm{pc}$)
are $\sim 1\, \mathrm{M_{\sun}/yr}$ and $\sim 10^6\,
\mathrm{M_{\sun}}$.  Since AGN lifetimes are much longer than 1 Myr,
mass must be transferred from larger radii to the inner disk. However,
at much larger radii this outer disk is gravitationally unstable and
must be forming stars. Thus, in order to feed the AGN, stability
arguments suggest that the rate of accretion must be greater than the
SFR. For simplicity we assume that $\Mdot_\rmn{BH} \propto
\Mdot_{\star}$. We inject energy into the ICM over a spherical region
of size $R_\rmn{AGN}$ about the AGN, according to
\begin{eqnarray}\label{eq:einj}
E_\rmn{inj} (<R_\rmn{AGN})&&=
\varepsilon_\rmn{r} \Mdot_{\star}(<R_\rmn{AGN}) c^2 \Delta t \\  && {\rm if} \  \Mdot_{\star}( <R_\rmn{AGN}) >
5\,\mathrm{M}_{\sun}/\rmn{yr}\, .\nonumber 
\end{eqnarray} 
The duty cycle over which the AGN outputs energy is $\Delta t$ and
$\varepsilon_\rmn{r}$ is an `efficiency parameter'. (As we describe
below, the calculated efficiency for turning mass into energy is much
smaller than $\varepsilon_\rmn{r}$.)  We have explored a wide range of
our two parameters, but the specific choices made for the figures are
$\Delta t=10^8$~yr and $\varepsilon_{\rmn{r}} =2\times 10^{-4}$. We require a
minimum SFR of $5\,\mathrm{M}_{\sun}/\rmn{yr}$ to activate AGN heating
in the halo it is housed in.

Given the output AGN energy, we must prescribe how it is to be
distributed. Our procedure is motivated by the way
\citet{2006MNRAS.366..397S} did AGN heating via bubbles. Using an
on-the-fly friends-of-friends (FOF) halo finding algorithm in
GADGET-2, we determine the mass and center of mass of each halo with
$M_\rmn{halo}>1.2\times 10^{12}\,h^{-1}\, \mathrm{M}_{\sun}$. We
calculate its global SFR within the AGN sphere of influence of radius
\begin{eqnarray}\label{eq:ragn}
R_\rmn{AGN}  = && \rmn{max} \left\{ \left[\frac{M_\rmn{halo}}{10^{15}\,h^{-1}\, \mathrm{M}_{\sun}}\right]^{1/3} [E(z)]^{-2/3}, \frac{u_\rmn{AGN}}{1 + z} \right\} \nonumber \\
&& \times 100\,h^{-1}\ {\rm kpc}, 
\end{eqnarray}   
where $u_\rmn{AGN} = \varepsilon_\rmn{s}$ and $E(z)^2 = \Omega_{\rmn{m}}(1+z)^3 + \Omega_\Lambda$.
Within the halos we partition $E_\rmn{inj}$ onto those gas particles
inside of $R_\rmn{AGN}$ according to their mass.  We have varied the
prescription for $R_\rmn{AGN}$ and its floor $u_\rmn{AGN}$ (chosen
here to be the gravitational softening $\varepsilon_\rmn{s}$); the
specific numbers given in eq.~\ref{eq:ragn} (and for
$\varepsilon_\rmn{r}$) match previous successful models that suppress
the over-cooling by means of AGN feedback \citep[][ see {\
}Sect.~\ref{sec:AGNtests}]{2008MNRAS.387.1403S}. Defining $R_{\Delta}$
as the radius at which the mean interior density equals $\Delta$ times
the critical density $\rho_\rmn{cr}(z)$ (e.g., for $\Delta =200$ or
500), then the ratio of $R_\rmn{AGN}$ to $R_{200}$ is a constant
$\sim 0.05$.

Although we have referred to our feedback mechanism as being caused by
AGN outflows, radiation pressure from stellar luminosity acting on
dust grains will serve much the same purpose, and could also deliver
high efficiencies \citep[e.g.][]{2005ApJ...630..167T}. In the code, we
have so far added $E_{\mathrm{inj}}$ as a pure heating component, but
it should allow for a mechanical, momentum-driven wind component as
well, which would not be as prone to catastrophic cooling and likely
decrease the $\varepsilon_{\rmn{r}} $ needed for useful star formation
suppression.

The relevant energy budget is not in fact defined by $\varepsilon_{\rmn{r}}$,
but rather by a redshift-dependent effective feedback efficiency
$\varepsilon_{\mathrm{eff}}\equiv \Sigma_i E_{\mathrm{inj},i} /
[M_{\mathrm{\star}}(<r)\,c^2]$, where we sum over every energy
injection event (labeled by $i$) and we calculate the stellar mass
$M_{\mathrm{\star}}(<r)$ within a given radius.  In all cases,
$\varepsilon_{\mathrm{eff}}\ll \varepsilon_{\rmn{r}}$, because: (i) heating
suppresses the stellar mass $\Delta M_{\star}$ created over $\Delta
t$, making it quite a bit less than the stellar mass $\Mdot_{\star}
\Delta t$ that would have formed without any feedback; and (ii)
$E_{\mathrm{inj}}$ is a stochastic variable, which we find to be zero
about half of the time because the required SF threshold is not
achieved. With our fixed $\varepsilon_{\rmn{r}} - R_\rmn{AGN}$ prescription,
our canonical g676 example has $\varepsilon_{\mathrm{eff}} \sim
5\times10^{-6}$ for the entire simulation; if all energy had been
released within the final $R_\rmn{AGN}$, $\varepsilon_{\mathrm{eff}} $
would be $8\times10^{-5}$, but feedback, especially at early times, is
much more widely distributed. Of a total $E_{\mathrm{inj}}=9\times
10^{61} \, {\rm ergs}$ for g676 we find $58\%$ is delivered in the
cluster formation phases at $z >2$, another $23\%$ is delivered in the
redshift range $1<z<2$ that can be probed with ACT and SPT resolution,
and only $19\%$ comes from the longer period below redshift
1. Feedback prescriptions with smaller $E_{\mathrm{inj}}$ which still
give the desired star formation suppression need further exploration.

\section{Pressure Profiles}

\begin{figure}
\epsscale{1.20}
\plotone{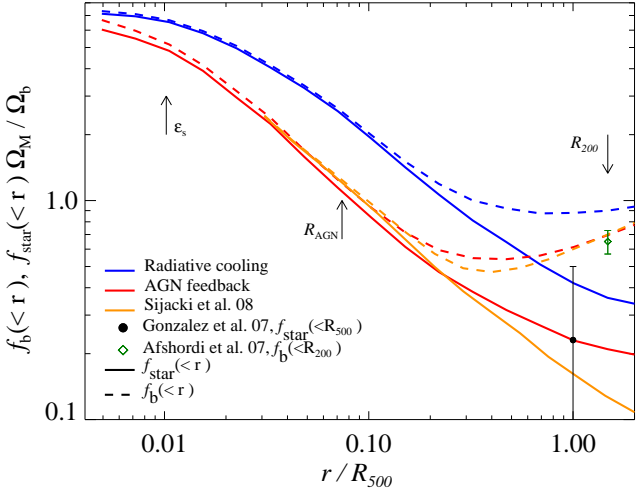}
\caption{Shown are $f_{\mathrm{b}}$ (dashed lines) and
  $f_{\mathrm{star}}$ (solid lines) normalized to the universal value
  ($f_{\mathrm{b}}=0.13$) assumed in our simulations of our cluster
  g676 with $M_{500} = 6.8 \times 10^{13}\,h^{-1}\,M_{\sun}$. The blue
  lines are for the simulation with radiative cooling and star
  formation while the red and orange lines are for our AGN feedback
  model ($\varepsilon_{\rmn{r}} =2\times 10^{-4}$, $\Mdot_{\star} \geqslant
  5\,\mathrm{M}_{\sun}/$yr) and that by \citet{2008MNRAS.387.1403S},
  respectively. The data points are observations by
  \citet{2007ApJ...666..147G} and
  \citet{2007MNRAS.378..293A}. $f_{\mathrm{star}} (<R_{500})$ from
  X-ray measurements also agrees well, but the errors are large. Our
  sub-grid model matches the results from \citet{2008MNRAS.387.1403S}
  in this high resolution simulation well.}
  \label{fig1}
\end{figure}

\subsection{Testing AGN feedback as resolution varies}
\label{sec:AGNtests}

AGN feedback self-regulates the star formation and energetics of a
cluster. In Fig.{\ }\ref{fig1} we compare the fraction of baryons
($f_{\mathrm{b}}$) and stars ($f_{\mathrm{star}}$) as functions of
cluster radius for the high-resolution `g676' simulations.  Our
radiative simulation produces $1.5-2$ times more stars than those with
AGN feedback.  Our sub-grid AGN model nicely reproduces the results in
\citet{2008MNRAS.387.1403S}.  It should also produce reliable results
in the cosmological box simulations in which over-cooling is less
severe because of the lower resolution. There is significant
sensitivity to the value chosen for the feedback parameter
$\varepsilon_{\rmn{r}}$: doubling it lowers $f_{\mathrm{b}}$ by a factor of
1.5, halving it increases $f_{\mathrm{star}}$ by 1.4. The
$100\,h^{-1}\,$ Mpc simulations were used to study the resolution
dependence of our feedback model by varying $N_{\mathrm{gas}}^{1/3}$
in steps from 64 to 256, with $\varepsilon_{\rmn{s}}$ and hence $u_\rmn{AGN}$
(eq.~\ref{eq:ragn}) decreased accordingly. As $u_\rmn{AGN}$ decreased,
$f_{\mathrm{star}}$ within $R_{500}$ increased almost linearly for
radiative cooling, whereas for AGN feedback the increases were
less. This can be traced to the hierarchical growth of structure since
in low-resolution simulations: the small star forming systems are
under-resolved; this decreases the SFR that mediates our AGN feedback;
and this lowers the overall number of stars produced in the
simulations. This behaviour is seen in other AGN feedback models
\citep{2007MNRAS.380..877S} and has been extensively studied in
non-AGN feedback simulations by \citet{2003MNRAS.339..312S}.

\subsection{Stacked pressure profiles}

\begin{figure}
\epsscale{1.2}
\plotone{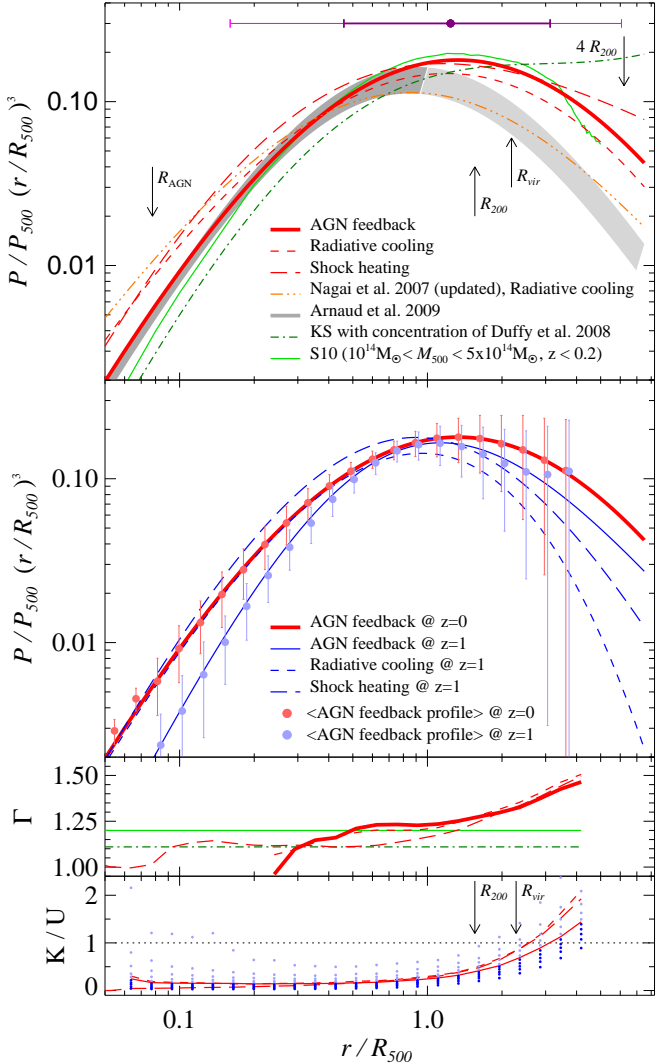}
\caption{{\em Top:} Comparison of fits to normalized average pressure
  profiles from analytic calculations, simulations and observations,
  scaled by $(r/R_{500})^3$. For a cluster of $M_{500} = 2 \times
  10^{14}\,h^{-1} M_{\sun}$, we show fits to our SPH simulations
  (red), and compare them with the analytic KS profile (green), the
  semi-analytic S10 average profile (light green), and a fit to AMR
  simulations \citep[updated profile by ][ private communication;
  orange]{2007ApJ...668....1N}. Our feedback model matches a fit to
  X-ray observations \citep[][grey bands]{2009arXiv0910.1234A} within
  $R_{500}$ well; only the dark grey part is actually a fit to the
  data, with the light grey their extrapolation using older theory
  results unrelated to the data.  We illustrate the 1 and 2 $\sigma$
  contributions to $Y_{\Delta}$ centered on the median for the
  feedback simulation by horizontal purple and pink error bars. {\em
  2nd panel:} We compare fits to our AGN model at redshift $z=0$ (red
  solid) to all our three models at redshift $z=1$ (blue). Shown are
  the 1$\sigma$ error bars of the cluster-by-cluster variance of the
  weighted averages in our AGN models using corresponding lighter
  colors. {\em 3rd panel:} We show the effective adiabatic index
  $\Gamma$ for our simulations, comparing it with KS (dash-dotted) and
  with a constant 1.2 (light green). {\em Bottom:} The distribution of
  kinetic-to-thermal energy in percentile decades is indicated by the
  dots for the feedback case, with the median shown for all three
  models; thus, there are significant additions to pressure support
  even in the cores of simulated clusters, and even more so in the
  SZ-significant outer parts.}
\label{fig2}
\end{figure}

For every halo identified by our FOF algorithm, we calculate the
center of mass, $R_{\Delta}$, the mass $M_{\Delta}$ within
$R_{\Delta}$ and compute the spherically-averaged pressure profile
normalized to $P_{\Delta} \equiv G M_{\Delta} \Delta\,
\rho_\rmn{cr}(z) f_{\mathrm{b}}/R_{\Delta}$, with $f_{\mathrm{b}} =
\Omega_{\mathrm{b}} / \Omega_{\mathrm{m}}$ \citep{2005RvMP...77..207V}
and radii scaled by $R_{\Delta}$. We then form a weighted average of
these profiles for the entire sample of clusters at a given
redshift. For Fig.{\ }\ref{fig2}, we have weighted by the integrated
y-parameter,
\begin{equation} 
Y_{\Delta} = {\sigma_{\rmn{T}}\over(m_\elct
c^2)}\int^{R_{\Delta}}_0 P_{\elct}(r) 4\pi r^2\, \dd r \,  \propto E_{\rmn{th}} (< R_{\Delta})\, , 
\end{equation}
where $\sigma_{\rmn{T}}$ is the Thompson cross-section, $m_\elct$ is
the electron mass and $P_{\elct}$ is electron pressure. For a fully
ionized medium the thermal pressure $P = P_{\elct} ({5X_{\rmn{H}} +
3}) / 2(X_{\rmn{H}} + 1) = 1.932 P_{\elct}$, where $X_{\rmn{H}} =
0.76$ is the primordial hydrogen mass fraction. Splitting the clusters
into a number of mass bins gives similar results to this monolithic
$Y_{\Delta}$ weight, as does weighting by $Y^2_{\Delta}$. We have
found that a simple parametrized model
\begin{equation}\label{eq:pfit} 
 P/P_{500} = A\left[1 +
\left(x/x_{\rmn{c}}\right)^{\alpha} \right]^{-\gamma/ \alpha}\, , \ x \equiv
r/R_{500}\, , \end{equation}
 with core-scale $x_{\rmn{c}}$,  amplitude $A$,  and two power law
indices, $\alpha$ and $\gamma$, fits better than with a fixed
$\alpha$. Sample values for our AGN feedback are $A = 82$, $x_{\rmn{c}} =
0.37$, $\alpha = 0.84$ and $\gamma = 4.6 $ at $z = 0$; generally the
parameters depend upon cluster mass and redshift. At $z \gtrsim 1$, a
more complex parameterization is needed.

In Fig.~\ref{fig2}, we show average pressure profiles multiplied by
$x^3$ to make them $\propto \dd E_{\rmn{th}}/\dd \ln r$, the thermal energy
per logarithmic interval in radius, and hence to $ \dd Y_{\Delta}/\dd
\ln r$. All profiles of $\dd E_{\rmn{th}}/\dd \ln r$ from simulations and
observations peak at or before $R_{200}$, but an integration to at
least $4 R_{200}$ is required for the total thermal energy to
converge. By contrast, the KS profile does not drop over this range
due to the constancy of $\Gamma$ and does not include the outer
cluster phenomena of asphericity, accretion shocks, {\it etc.}
Throughout this paper, we have computed the KS model with an updated
concentration parameter given by \citet{2008MNRAS.390L..64D}. We also
show a scaled average S10 pressure profile for clusters with $10^{14}
M_{\sun} < M_{500} < 5 \times 10^{14}$ and redshift $< 0.2$. The S10
profile has been weighted by $Y_{\Delta}$ and agrees well within
$R_{500}$ and with a slight excess pressure beyond $R_{500}$.
 
 Fig.~\ref{fig2} shows our feedback model traces the observed
"universal" X-ray profile of \citet{2009arXiv0910.1234A} shown as a
dark-grey band rather well within $R_{500}$. This fit came out
naturally, with no further tuning of our feedback parameters beyond
trying to agree with the \citet{2008MNRAS.387.1403S} simulation. Our
models without AGN feedback have larger pressures inside
$R_{500}$. For the light grey band beyond $R_{500}$, the universal
X-ray profile did not use observations, but was fit to an average
profile of earlier simulations so the deviation $> R_{500}$ does not
represent a conflict of our profiles with the data, rather with the
earlier simulations.  The band shown for the X-ray profile gives a
crude correction for the bias in $M_{500}$ and $R_{500}$ resulting
from the \citet{2009arXiv0910.1234A} assumption of hydrostatic
equilibrium. This yields mass values which are on average 25\% too low
\citep{2007ApJ...668....1N}, so the band represents a 0-25\%
uncertainty in $M_{500}$. This change only affects $R_{500}\propto
M_{500}^{1/3}$ and $P_{500} R_{500}^3 \propto M_{500}^{5/3}$ but does
not affect the shape of the profile. (However, as the bottom panel
shows, such a correction from turbulence and un-virialized bulk motions
\citep{2006ApJ...650..128K} will depend upon radius and selection
function of the X-ray clusters used to make the fit.)

Another important issue is the
relation between the $Y_{\Delta}$ and cluster mass. We fit our results
for this to the scaling relation 
\begin{equation}\label{eq:ym}
Y_{500} = 10^B \, (M_{500}/3\times 10^{14}\, \hsev^{-1}\,M_{\sun})^A \, \hsev^{-5/2} \, \rmn{Mpc}^2,
\end{equation} 
where $\hsev \equiv 0.7 \times 100 \mbox{ km s}^{-1} \mbox{
Mpc}^{-1}$.  Parameters from our simulation are $B=(-4.47 \pm
0.08,-4.46 \pm 0.20,-4.5 \pm 0.1)$ and $A=(1.66 \pm 0.12,1.71 \pm
0.25,1.75 \pm 0.06)$ for the sequence (1) shock heating, (2) radiative
cooling and (3) AGN feedback.  These values are similar to the
$B=-4.739 \pm 0.003$ and $A = 1.790 \pm 0.015$ found by
\citet{2009arXiv0910.1234A}, as well as the $B=-4.713 \pm 0.004$ and
$A = 1.668 \pm 0.009$ found by S10. We note that
\citet{2009arXiv0910.1234A} actually used a mass proxy in place of
$M_{500}$, so their errors are not representative of the true
observational scatter in the $Y-M$ scaling relation. The AGN feedback
model of \citet{2007MNRAS.380..877S} was also able to reconcile the
cluster X-ray luminosity and temperature scaling relation
\citep{2008ApJ...687L..53P}.

We find a large variation in the outer pressure profiles beyond
$R_\rmn{vir}$, especially at redshift $z\sim1$ as is shown in the
second panel of Fig.~\ref{fig2}. These regions may have sub-halos, and
external but nearby groups on filaments, most of which will eventually
be drawn into the clusters. In spite of the large variance of the
scaled profiles, the fit to the profiles at $z=0$ follows the
average. At larger redshift, however, our fitting formula will require
more degrees of freedom than in eq.~\ref{eq:ym} to reflect the range
of behaviour of the highly dynamical outer regions.

\begin{figure}
\epsscale{1.2}
\plotone{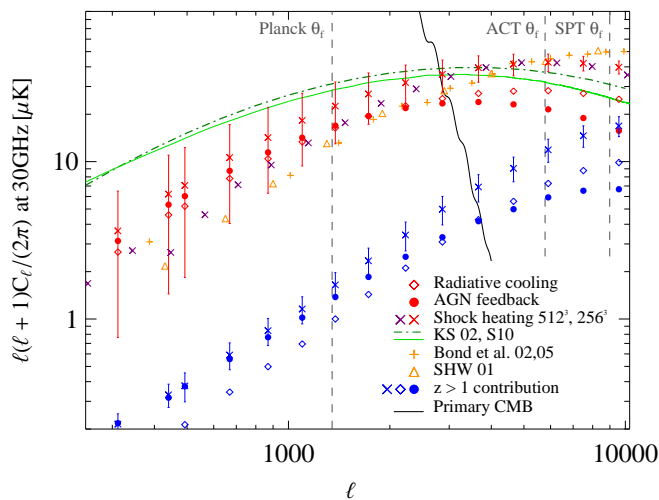}
\caption{Predictions for the tSZ power spectrum at 30 GHz from our simulations
  (red and purple symbols), simulations by \citet{2001ApJ...549..681S} (orange
  triangles), simulations by \citet{2005ApJ...626...12B} (orange pluses),
  semi-analytical simulations by S10 (dark green) and analytical calculations by
  KS (light green). The 256$^3$ power spectra (red symbols) are averages over 12
  translate-rotate tSZ maps and 10 separate hydrodynamical
  simulations for each of the 33 redshift bins, the power spectra of which are
  then added up to yield the total spectrum; the error bars show the variance among
  the power in all maps. The full-width half-max values appropriate for Planck,
  ACT and SPT show which part of the templates these experiments are sensitive
  to.  At low-$\ell$, the discrepant higher power in the semi-analytical
  calculations can be traced to the enhanced pressure structures assumed beyond
  $R_{200}$ over what we find.}
\label{fig3}
\end{figure}

\section{SZ Power Spectra from Hydrodynamical Simulations}

\subsection{Stacked SZ power spectra of translated-rotated cosmological boxes}

We randomly rotate and translate our simulation snapshots at different
redshifts \citep[][B0205]{2000MNRAS.317...37D, 2001ApJ...549..681S}.
To obtain thermal Compton-$y$ maps, we perform a line-of-sight
integration of the electron pressure within a given solid angle,
i.e. $y = \sigma_{\rmn{T}}\int n_\elct k\,T_\elct/(m_\elct c^2)\, \dd l$,
where $k$ is the Boltzmann constant, $n_\elct$ and $T_\elct$ are the
number density and temperature, respectively. We construct $1.6$\degr$
\times 1.6$\degr and $3.2$\degr$ \times 3.2$\degr maps for the $256^3$
and $512^3$ simulations, respectively. Using this method there are
large sample variances \citep{2002ApJ...579...16W} associated with
nearby cluster contamination. We have quantified their influence on
the power spectrum for each of our three physics models by averaging
over twelve translate-rotate viewing angles each projected from our
ten $256^3$ full hydrodynamical simulations for each of the 33
redshift outputs back to a redshift $z=5$; the power spectra of which
are then added up to yield the total spectrum. This method of
computing the power spectrum has the advantage of taking care of the
artificial correlations that occur because any individual simulation
follows the time evolution of the same structure.  For the shock
heating case, we did ten more hydrodynamical simulations to show that
our averaged template had converged (within $\sim$10\%), but note that
using only a few boxes can be misleading in terms of rare events. 

The computationally more expensive $512^3$ SZ spectra have the
equivalent of 8 $256^3$ plus wider coverage, so the $512^3$ shock
heating result shown gives a reasonable indication of what to expect.
The other 2 physics single-box cases at $512^3$ are similar to the
$256^3$ ensemble means. The analytical approach has the great
advantage of including an accurate mean cluster density to high halo
masses, but to be usable for SZ power estimation, scaled pressure
profiles must also be accurate, a subject we turn to in future
work. For now, we note that using such profiles from our simulations
gives good agreement with the average SZ power shown at the low $\ell$
where sample variance will be largest.  In Fig.~\ref{fig3}, our
simulation templates and the KS template shown have excluded
structures below $z = 0.07$ to decrease the large sample variance
associated with whether a large-ish cluster enters the
field-of-view. Such entities would typically be removed from CMB
fields and considered separately.

The mean Compton $y$-parameter found
in our AGN feedback simulations is one order of magnitude below the
COBE FIRAS upper limit of $15 \times 10^{-6}$
\citep{1996ApJ...473..576F}.

We compare the theoretical predictions for the tSZ power spectrum in
Fig.~\ref{fig3}. Our $512^3$ and $256^3$ shock heating simulations are
in agreement with previous SPH simulation power spectra
\citep[][B0205]{2001ApJ...549..681S} scaled by $C_\ell \propto
(\Omega_\rmn{b} h)^2 \Omega_\rmn{m} \sigma_8^{7}$, with the factors
determined from our simulations of differing cosmologies. The B0205 SZ
power shown had a cut at $z = 0.2$, appropriate for CBI fields; using 
the same cut on a shock heating simulation with the same
cosmology that we have done, we get superb agreement. 

The KS and S10 semi-analytic SZ power spectra templates differ
substantially from our templates, in particular with higher power at
low $\ell$: as shown in Fig.~\ref{fig2}, the KS pressure profile
beyond $R_{500}$ overestimates the pressure relative to both
simulations and observations, leading to the modified shape and larger
$Y_{\Delta}$; this behaviour is also shown in
\citet{2010arXiv1001.4538K}. The spectrum from S10 is very similar to
KS possibly because both assume hydrostatic equilibrium, and a
polytropic equation of state with a fixed adiabatic index, $\Gamma
\sim 1.1-1.2$.  Inside $R_{200}$, these assumptions are approximately
correct, but they start to fail beyond $R_{200}$. A demonstration of
this is the rising of $\Gamma$ and of the ratio of kinetic-to-thermal
energy $K/U$ shown for our simulations in the bottom panels of
Fig.~\ref{fig3}. The present day ($a=1$) internal kinetic energy of a
cluster is given by $K \equiv \Sigma_i m_{\rmn{gas},i} \left|\bvel_i -
\bar{\bvel}+ H_0(\vecbf{x}_i-\bar{\vecbf{x}})\right|^2 /2$, where
$H_0$ is the present day Hubble constant, $\bvel_i$ and $\vecbf{x}_i$
are the peculiar velocity and comoving position for particle $i$, and
$\bar{\bvel}$ and $\bar{\vecbf{x}}$ are the gas-particle-averaged bulk
flow and center of mass of the cluster. The additional thermal
pressure support we find at large radii from AGN feedback results in
the slightly slower rate of $K/U$ growth shown. In all cases the large
kinetic contribution shown should be properly treated in future
semi-analytic models.

Varying the physics over the three cases for energy injection in our
simulations leads to relatively minor differences in Fig.~\ref{fig3}
among the power spectra for $\ell \lesssim 2000$. This agreement is
due in part to hydrostatic readjustment of the structure so the virial
relation holds, which relates the thermal content, hence $Y_{\Delta}$,
to the gravitational energy, which is dominated by the dark matter.
Our AGN feedback parameters do not lead to dramatic gas expulsions to
upset this simple reasoning. Our radiative cooling template has less
power at all scales compared to the shock heating template since
baryons are converted into stars predominantly at the cluster centers
and the ICM adjusts adiabatically to this change. Thus, at low $\ell$
where clusters are unresolved, shock heating and radiative simulations
give upper and lower limits, bracketing the AGN feedback case. AGN
feedback suppresses the core value of the pressure compared to the
radiative simulation resulting in less power at $\ell > 2000$, a trend
that is more pronounced at $z>1$ (as shown in Fig.~\ref{fig3}). Thus,
at these angular scales, the power spectrum probes the shape of the
average pressure profile. It depends sensitively on the physics of
star and galaxy formation e.g., \citet{2008ApJ...678..674S}. Over the
$\ell$-range covered by Planck, these effects are sub-dominant, and
serve to highlight the importance of the high-resolution reached by
ACT and SPT.

\begin{figure}
\epsscale{1.2}
\plotone{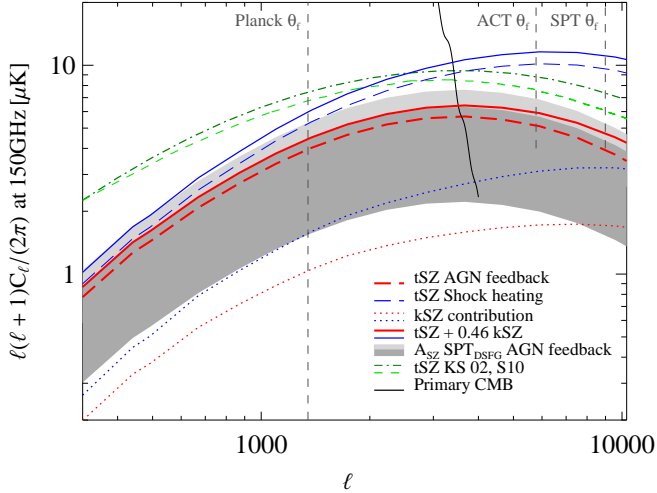}
\caption{Our 150 GHz tSZ adiabatic and feedback ($A_{\mathrm{SZ}}=1$)
power spectra computed with $\sigma_8 = 0.8$ (long dashed lines) are
contrasted with the dark grey band indicating the $1\sigma$ range in
multiplicative amplitude, $A_{\mathrm{SZ}}$ $=0.75 \pm 0.36$, allowed
by the SPT$_{\rmn{DSFG}}$ power spectrum for the feedback template
shape. The light grey band is the $2\sigma$ upper limit region. The
$A_{\mathrm{SZ}}=1$ S10 tSZ power spectrum (dashed line) and the KS
tSZ spectrum (dash dotted line) are shown for contrast; their allowed
$1\sigma$ band is determined by multiplying these by their
$A_{\mathrm{SZ}}$ values given in Table~1, but cover a similar swath
to the grey bands.  We also show the averaged kSZ power spectra
computed for our simulations by dotted lines.  The kSZ spectra were
calculated in the same was as the tSZ spectra were, and have similar
shapes.  However, kSZ is underestimated at low $\ell$ because of
missing bulk velocities in the simulations. There should be an
additional (rather uncertain) kSZ template from inhomogeneous
re-ionization as well.  To show the tension with the CMB data, we plot
the tSZ + 0.46 kSZ power (solid lines) since this can be directly
compared with the SPT$_{\rmn{DSFG}}$ grey bands.  }

\label{fig4}
\end{figure}

\subsection{Current constraints on SZ template amplitudes and $\sigma_\rmn{8,SZ}$}

Instead of varying all cosmological parameters on which the thermal
and kinetic SZ power spectra, $C_{\ell,\mathrm{tSZ}}$ and
$C_{\ell,\mathrm{kSZ}}$, depend, we freeze the shapes by adopting the
parameters for our fiducial $\sigma_8=0.8$ (and $\Omega_{\rmn{b}}h=0.03096$)
model evaluated at 150 GHz, and content ourselves with determining
template amplitudes, $A_{\mathrm{tSZ}}$ and $A_{\mathrm{kSZ}}$, and a
total SZ amplitude $A_{\mathrm{SZ}}$:
\begin{equation}\label{eq:clsz}
A_{\mathrm{SZ}}C_{\ell,\mathrm{SZ}} \equiv 
f(\nu)A_{\mathrm{tSZ}}C_{\ell,\mathrm{tSZ}} +
A_{\mathrm{kSZ}}C_{\ell,\mathrm{kSZ}}\, . 
\end{equation}
The spectral function for the tSZ \citep{1970Ap&SS...7....3S},
$f(\nu)$, vanishes at the SZ null at $\sim$ 220 GHz and we normalize
it to unity at $\nu =$ 150 GHz, so it rises to ~4 at 30 GHz. Therefore
if we find values of $A_{\mathrm{SZ}}$ below unity then either
$\sigma_{8}$ is smaller than the fiducial cosmological value as
derived from the primary CMB anisotropies, or else the theoretical
templates overestimate the SZ signal.

To determine the probability distributions of these amplitudes and
other cosmological parameters from current CMB data we adopt Markov
Chain Monte Carlo (MCMC) techniques using a modified version of
CosmoMC \citep{2002PhRvD..66j3511L}. We include WMAP7
\citep{2010arXiv1001.4635L} and, separately, ACT
\citep{2010arXiv1001.2934T} and SPT \citep{2009arXiv0912.4317L}.  In
all cases, we assume spatial flatness and fit for 6 basic cosmological
parameters ($\Omega_{\rmn{b}}h^2$, $\Omega_{\rmn{DM}}h^2$, $n_{\rmn{s}}$, the primordial
scalar power spectrum amplitude $A_{\rmn{s}}$, the Compton depth to
re-ionization $\tau$, and the angular parameter characterizing the
sound crossing distance at recombination $\theta$).  We also allow for
a flat white noise template $C_{\ell,\mathrm{src}}$ with amplitude
$A_{\mathrm{src}}$, such as would arise from populations of unresolved
point sources. We marginalize over $A_{\mathrm{src}}$, allowing for
arbitrary (positive) values. Generally there will also be a spatial
clustering component for such sources, and these will have templates
that are partially degenerate in shape with that for tSZ, but because
of the large uncertainties we ignore such contributions here. Reducing
the SZ and unresolved source problems to determinations of overall
amplitudes multiplying shapes has a long history, e.g., the CBI
sequence of papers, and was adopted as well by the ACT and SPT
teams. Our results differ slightly from those reported by the ACT team
because they use WMAP5+ACT and a combined tSZ+kSZ S10-template, and by
the SPT team who use WMAP5+QUaD+ACBAR+SPT and add constraints on the
white noise source amplitude beyond the non-negativity we impose.

\begin{table*}
  \caption{Constraints on $A_{\mathrm{SZ}}$ and $\sigma_\rmn{8,SZ}$}
  \label{tab:sig8}
 \begin{center}
    \leavevmode
    \begin{tabular}{llllllll} \hline \hline              
   tSZ template   &  \multicolumn{2}{c}{ACT 148 GHz} & \multicolumn{2}{c}{SPT 153 GHz} & 
   \multicolumn{2}{c}{SPT$_{\rmn{DSFG}}$} \\ \hline
   & $A_{\mathrm{SZ}}$& $A_{\mathrm{tSZ}}$ &$A_{\mathrm{SZ}}$& $A_{\mathrm{tSZ}}$& $A_{\mathrm{SZ}}$& $A_{\mathrm{tSZ}}$ \\ \hline
 KS                & $<$ 1.55 & $<$ 1.26 & $1.01 \pm 0.25$ & $0.72 \pm 0.25$ & $0.43 \pm 0.21$ & $ 0.30 \pm 0.21$ \\
 S10               & $<$ 1.95 & $<$ 1.67 & $1.39 \pm 0.34$ & $1.11 \pm 0.34$ & $0.50 \pm 0.25$ & $ 0.38 \pm 0.25$ \\
 Shock heating     & $<$ 2.13 & $<$ 1.84 & $1.13 \pm 0.28$ & $0.84 \pm 0.28$ & $0.44 \pm 0.22$ & $ 0.31 \pm 0.22$ \\
 Radiative cooling & $<$ 2.75 & $<$ 2.46 & $1.50 \pm 0.37$ & $1.21 \pm 0.37$ & $0.59 \pm 0.29$ & $ 0.45 \pm 0.29$ \\
 Feedback          & $<$ 2.93 & $<$ 2.66 & $1.76 \pm 0.43$ & $1.49 \pm 0.43$ & $0.75 \pm 0.36$ & $ 0.63 \pm 0.36$ \\\hline
& $\sigma_\rmn{8,SZ}$ & $\sigma_\rmn{8,tSZ}$ & $\sigma_\rmn{8,SZ}$ & $\sigma_\rmn{8,tSZ}$ & $\sigma_\rmn{8,SZ}$ & $\sigma_\rmn{8,tSZ}$ \\ \hline 
 KS                & $<$ 0.864 & $<$ 0.845 & $0.792^{+0.029}_{-0.029}$ & $0.757^{+0.039}_{-0.037}$ & $0.690^{+0.057}_{-0.055}$ & $ 0.622^{+0.105}_{-0.051}$ \\
 S10               & $<$ 0.891 & $<$ 0.874 & $0.828^{+0.031}_{-0.030}$ &$ 0.800^{+0.038}_{-0.036}$ & $0.705^{+0.060}_{-0.058}$ & $ 0.636^{+0.109}_{-0.054}$ \\
 Shock heating     & $<$ 0.900 & $<$ 0.883 & $0.804^{+0.031}_{-0.030}$ &$ 0.768^{+0.040}_{-0.038}$ & $0.691^{+0.059}_{-0.057}$ & $ 0.607^{+0.119}_{-0.057}$ \\
 Radiative cooling & $<$ 0.935 & $<$ 0.922 & $0.837^{+0.032}_{-0.031}$ &$ 0.809^{+0.039}_{-0.037}$ & $0.721^{+0.060}_{-0.058}$ & $ 0.660^{+0.102}_{-0.053}$ \\
 Feedback          & $<$ 0.944 & $<$ 0.932 & $0.856^{+0.033}_{-0.032}$ &$ 0.835^{+0.038}_{-0.037}$ & $0.746^{+0.061}_{-0.059}$ & $ 0.703^{+0.091}_{-0.055}$ \\\hline
    \end{tabular}
    \begin{quote}
      \noindent The mean and standard deviation of the thermal SZ
       power spectrum template amplitude $A_{\rmn{tSZ}}$ and the total
       SZ, including our computed kSZ contribution. The numbers assume
       the kSZ template is perfectly degenerate in shape with the tSZ
       one. $A_{\rmn{SZ}}=A_{\rmn{tSZ}}+A_{\rmn{kSZ}}$ at 150 GHz,
       with the relative enhancement in our simulations given by
       $A_{\mathrm{kSZ}}/A_{\mathrm{tSZ}}$ $= 0.29, 0.29, 0.27$ for
       the shock heating, radiative cooling and feedback simulations,
       respectively. We have used the ACT team's 148 GHz power
       spectrum, the SPT team's 153 GHz spectrum and the SPT
       DSFG-subtracted (SPT$_{DSFG}$) spectrum, along with WMAP7. The
       amplitude of the SZ power is normalized to our fiducial
       $\sigma_8 = 0.8$ cosmology. A rough guide to the $\sigma_8$
       tension is obtained in the lower rows, using $\sigma_\rmn{8,SZ}
       \propto A_{\mathrm{SZ}}^{1/7}\left ( \Omega_{\rmn{b}} h\right
       )^{-2/7}$, with exponents determined by B0205 and KS. Since kSZ
       varies more slowly with $\sigma_8$ than tSZ, the numbers are
       just indicative.
    \end{quote}
  \end{center}
\end{table*}

We first consider a simplified case with $A_{\mathrm{kSZ}} $
constrained to be zero and all other cosmic parameters and the source
amplitude marginalized, yielding a probability distribution for
$\asz$. The means and standard deviations from our MCMC runs are given
in the upper rows of Table \ref{tab:sig8} in columns 2, 4 and 6 for a
number of data combinations and for our 3 physics simulation cases,
contrasting with KS and S10. The ACT data is for 148 GHz. There are
two SPT cases given. The first uses just the 153 GHz spectrum so it
can be directly compared to ACT.  For SPT, \citet{2009arXiv0912.4317L}
also report a power spectrum derived from subtracting a fraction $x$
of their 220 GHz data from the 153 GHz data to minimize the
contribution from dusty star-forming galaxies (DSFG); since 220 GHz is
the SZ null, this does not modify the tSZ contribution, but would
diminish the frequency-flat kSZ.  However, a normalization factor is
chosen to preserve power for primary CMB signals that are flat in
frequency like kSZ.  This has the effect of boosting the tSZ power by
a factor of $(1-x)^{-2}$.  \citet{2009arXiv0912.4317L} find that
$x=0.325$ minimizes the contribution from the DSFGs so the
DSFG-subtracted spectrum suppresses the kSZ by a factor of 0.46
relative to the tSZ. A $\sim$25\% uncertainty remains in $x$ which
should be taken into account statistically, but is not here. The
correct approach would be to simultaneously treat the 153 GHz and 220
GHz cases, with full modelling of the different classes of point
sources, including their clustering, and to take into account the
non-Gaussian nature of the SZ and source signals which impact sample
variance.

The ACT data is only giving upper limits with their current published
data, whereas SPT has detections at 153 GHz with $\asz$ compatible
with unity.  For the SPT 153 GHz-only spectrum, we find S10 gives
$\asz=1.39 \pm 0.34$ while the feedback template gives $\asz=1.76 \pm
0.43$, and the comparable 95\% upper limits from ACT are 1.95 and
2.93. However, although the white noise shape has been vetoed by
marginalization, there could be a residual clustered source
contribution from dusty galaxies pushing the derived $\asz$ high. To
the extent that SPT$_{DSFG}$ vetoes this DSFG clustering as well as
their Poisson contribution, that $\asz$ would be a better
indicator. It shifts from $0.43 \pm 0.21$ for KS and $0.50 \pm 0.25$
for S10 up to $0.75 \pm 0.36$ for the feedback template, an increase
of 50\%.  The large difference between the 150 and source-subtracted
templates, even after marginalizing over a Poisson term, may suggest
the power in the correlated source component may be similar to the SZ
power, emphasizing the work necessary to do a correct treatment.

Any non-zero kSZ contribution will take some of the amplitude from
$\asz$, leaving even smaller $\atsz$ values; columns 3, 5 and 7 of the
table give estimates of this diminution.  The kSZ power spectra that
we have computed are broadly similar to the tSZ power shape, with
however sufficiently significant differences to allow shape
discrimination in addition to the frequency separability, as
Fig.~\ref{fig4} shows.  At 150 GHz and an $\ell = 3000$ pivot, we find
the kSZ power is $\sim 29$\%, $\sim 29$\% and $\sim 27$\% of the tSZ
power for the shock heating, radiative cooling and feedback
simulations, respectively. We normalize the kSZ to the tSZ at this
pivot of 3000 since it has most of the constraining power in the
CosmoMC chains for the ACT and SPT measurements and results in the
smallest error bars: on larger scales, the errors are increased by the
contribution from primary anisotropies while smaller scales are
dominated by the instrumental and galaxy-source shot noise.

We used exactly the same procedure to obtain the kSZ spectrum as we
used for the tSZ spectrum.  The temperature decrement due to the kSZ
effect is $\Delta T/T = \sigma_\rmn{T} \int n_{\mathrm{e}}\,
\vel_{\rmn{r}}/c\, \dd l$, where $\vel_{\rmn{r}}$ is the radial peculiar velocity
of the gas relative to the observer. We constructed 12
translate-rotate kSZ maps for each of our 10 separate hydrodynamical
simulations and for each of the 41 redshift bins back to $z=10$
(rather than $z=5$ for tSZ), computing the average and variance of all
of these.  Since we use simulations with side length
$L=165\,h^{-1}$Mpc for our $256^3$ cases, with fundamental wavenumber
$(26\,h^{-1}{\rm Mpc})^{-1}$, our spectra are missing a bit of power
on the largest scales (affecting low-$\ell$) since we do not sample
well the long-wavelength tail of the velocity power spectrum in spite
of the number of runs done.

We have included the kSZ template by ignoring the relatively small
shape difference about the pivot point of the kSZ and tSZ power
spectra; i.e., we assume the perfect degeneracy $C_{\ell,\mathrm{kSZ}}
= C_{\ell,\mathrm{tSZ}}$, as the SPT team did. Thus we only need the
ratios $A_{\mathrm{kSZ}}/A_{\mathrm{tSZ}}$ given above for the 150 GHz
cases and the further $x$ factors for the mixed frequency DSFG
case. For the ratios we use our translate-rotate values of 0.29, 0.29
and 0.27 from our simulations, 0.276 for S10, and used a rough
estimate of 0.25 for KS. Apart from ignoring the shape difference, we
have also ignored kSZ from patchy re-ionization at high redshift,
although it can have a competitive amplitude to the late time fully
ionized gas motions with respect to the CMB rest frame that we are
modelling \citep{2008MNRAS.384..863I, 2007ApJ...660..933I}.  In
presenting the results from our analyses of the MC Markov chains, we
just subtract $\aksz$ from $\asz$. The Table~\ref{tab:sig8} $\atsz$
that we derive from these assumptions are all on the low side of unity
for DSFG, with KS and S10 being more than $2.5\sigma$ low, whereas the
feedback template is only about $1\sigma$ low (and $1\sigma$ high for
153 GHz alone). We leave it to future work to include a more complete
implementation of the kSZ spectra.

The means and errors on $A_{\mathrm{SZ}}$ provide the cleanest way of
presenting the tension, or lack thereof, of these SZ models with the
primary CMB data which indicates $\sigma_8 \approx 0.8$. However, it
has been conventional to translate these numbers into a
$\sigma_\rmn{8,SZ} $ using the way $A_{\mathrm{SZ}}$ scales with
cosmic parameters, roughly as $A_{\mathrm{SZ}} \propto \sigma_8^7\left
(\Omega_{\rmn{b}}h\right )^2$, as given by B0205 and KS. The lower rows in
Table~1 show $\sigma_\rmn{8,SZ} $ using this scaling. Although the
scaling applies to the tSZ component only, with the kSZ power being
less sensitive to $\sigma_8$, we also quote results for the
kSZ-corrected cases. Ideally one should use the data to determine the
cosmic parameters which uniquely and fully determine the primary
spectrum, the $A_{\mathrm{tSZ}}$ and $A_{\mathrm{kSZ}}$, and the tSZ
and kSZ shape modifications as the parameters vary. This slaved
treatment enforcing $\sigma_\rmn{8,SZ} =\sigma_\rmn{8} $ has
$\sigma_\rmn{8} $'s value being driven by WMAP7 and other primary CMB
data rather than by the SZ information.

\section{Conclusions and Outlook}

Without hydrodynamical simulations in a cosmological framework similar
to the ones presented in this paper it is hard to come up with a
consistent model of the gas distribution in clusters and the infall
regions which both contribute significantly to the SZ power
spectrum. In this paper, we identify three main points that a future
semi-analytic model of such a pressure distribution has to provide.

(1) In order to arrive at a consistent gas distribution that matches
not only the integrated stellar mass fraction but also the X-ray
derived pressure profiles within $R_{500}$, we need self-regulating
AGN-type feedback. We emphasize that we tuned our parameters to match
a previous single-cluster model that successfully suppressed the
over-cooling by means of AGN feedback \citep{2008MNRAS.387.1403S}. The
excellent agreement with current data was a pleasant byproduct: our
simulated pressure profiles agree with recently obtained observational
ones that have been constructed from X-ray data; the scaling relations
between the cluster mass and X-ray based Compton-$Y$
\citep{2009arXiv0910.1234A} also agree; as do the integrated stellar
and gas mass fractions
\citep{2007ApJ...666..147G,2007MNRAS.378..293A}.

(2) The amount of non-gravitational energy injection into
proto-clusters and groups by AGN and starburst galaxies at
intermediate-to-high redshifts $z\gtrsim0.8$ is poorly
understood. Other observables are needed to constrain the physics and
to answer this question which seems to be essential in understanding
the resulting gas profiles. Our simulations suggest that AGN-type
feedback lowers the central pressure values as a hydrodynamic response
of the gas distribution to the non-gravitational feedback of
energy. This effect inhibits gas from falling into the core
regions which causes a flatter and more extended pressure profile and
a noticeably reduced power of the SZ power spectrum at small angular
scales for $\ell\gtrsim2000$.

(3) For the SZ flux to be converged, an integration of the pressure
profile out to $4 R_{200}$ is necessary; half of the SZ flux is
contributed from regions outside $R_{200}$. To compute a reliable SZ
power spectrum, it is essential to precisely characterize the state of
the gas in these infall regions. In particular, we find that: (i) the
pressure support from kinetic energy strongly increases as a function
of radius to reach on average equipartition with the thermal energy at
$\sim 2 R_{200}$ in our AGN model with the exact dependence on cluster
mass to be determined by future work; (ii) the effective adiabatic
index $\Gamma =\dd\ln p/\dd\ln \rho \sim 1.2$ in the interior, but
upturns towards $\Gamma \sim 5/3$ beyond the virial radius; (iii) the
inclusion of cluster asphericity at large radii may also become
important.

Hence a successful semi-analytic model of the spherical cluster
pressure, if that is indeed a viable goal, at the least needs careful
calibration using numerical simulations which accurately treat all of
the effects. The variance of the average profiles also encodes
important information that is manifested in the power spectrum. Our
studies also show that simplified analytic models that employ
hydrostatic gas models with a constant $\Gamma$ necessarily
overpredict the SZ power on large scales by up to a factor of two and
predict an inconsistent shape of the SZ power spectrum. The
alternative that we explore in a subsequent paper is to use stacked
scaled simulational clusters which are rotated to principal axes to provide the
pressure form factors for the semi-analytic approach.

The tSZ power spectrum of our $512^3$ simulation agrees well with the
average of our ten $256^3$ simulations. A large number of simulations
are needed to properly sample the high-mass end of the cluster mass
function and hence accurately deal with sample (cosmic)
variance. Alternatively, larger cosmological volumes can compensate
since they contain enough statistics on the large scale modes that are
responsible in part for forming the highest-mass clusters which are
also the rarest events. This, however, is quite challenging as we
require the same (high-)resolution to accurately follow the physics in
the cluster cores which is needed to obtain profiles that match
current X-ray data. Our $256^3$ simulations do not quite sample large
enough scales to provide a fully converged kSZ power spectrum at low
$\ell$ since we miss the long-wavelength tail of the velocity power
spectrum. We also have ignored the patchy re-ionization kSZ which could
be a significant contributor, up to $50\%$ of the total kSZ
\citep[e.g.,][]{2008MNRAS.384..863I, 2007ApJ...660..933I}.
  
 We have found the $\ell <2000$ multipole range to be relatively
insensitive to cooling and feedback, at least for the range
constrained by the X-ray data. We did find the higher multipole range
($\ell\sim2000-10000$) probed by the high-resolution ACT and SPT CMB
telescopes is sensitive to the feedback prescription; hence the
high-$\ell$ SZ power spectrum can be used to constrain the theory of
intracluster gas, in particular for the highly uncertain redshifts
$>0.8$.  In addition to the SZ power spectrum probe, our simulations
can be used to address the cosmological significance of cluster counts
as derived from the SZ effect. Counts provide complementary
constraints on parameters that help to break some degeneracies that
are present in the power spectrum method. By employing inhomogeneous,
localized and self-regulated feedback we are not only able to match
recent X-ray reconstructions of cluster core regions, but also
decrease the tension in $\sigma_8$ estimated from SZ power with
$\sigma_8$ from other cosmological probes. However, only a detailed
confrontation between simulations exploring the vast terrain of
feedback options with the rapidly improving high resolution
observations of cluster interiors can move the theory of cluster gas
physics and its use for precision cosmology forward.

%\acknowledgments

We thank Norm~Murray, Volker~Springel, Hy~Trac, Jerry~Ostriker,
Gil~Holder, Niayesh Afshordi and Diasuke Nagai for useful
discussions. Research in Canada is supported by NSERC and
CIFAR. Simulations were run on SCINET and CITA's Sunnyvale HPC
clusters.

% --- section: bibliography --- %
\bibliography{bibtex/chp}
\bibliographystyle{apj}

\clearpage

\end{document}